\documentclass[aps,pra,10pt,amsmath,onecolumn,notitlepage,tightenlines,superscriptaddress,floatfix,eqsecnum]{revtex4-1}
\usepackage{amssymb}
\usepackage{bbold}
\usepackage{upgreek}

\newcommand{\ket}[1]{|#1\rangle}
\newcommand{\bra}[1]{\langle#1|}
\newcommand{\braket}[2]{\langle#1|#2\rangle}
\newcommand{\proj}[1]{\ket{#1}\bra{#1}}

\newcommand{\norm}[1]{\left| #1 \right|}

\def\sB{\mathcal{B}}
\def\sC{\mathcal{C}}
\def\sE{\mathcal{E}}
\def\sF{\mathcal{F}}
\def\sH{\mathcal{H}}
\def\sI{\mathcal{I}}

\def\sO{\mathcal{O}}

\def\bI{{\bf I}}

\newcommand{\m}{\hspace{0.1pt}}
\newcommand{\n}{\hspace{-1.0pt}}
\newcommand{\req}[1]{Eq.~(\ref{#1})}

\DeclareMathOperator{\tr}{tr}

\begin{document}

\title{\bf Ancilla models for quantum operations:\\
For what unitaries does the ancilla state have to be physical?}
\date{March 2012}
\author{Zhang Jiang}
\affiliation{Center for Quantum Information and Control, University of New Mexico, MSC07-4220, Albuquerque, New Mexico 87131-0001, USA}
\author{Marco Piani}
\affiliation{Institute for Quantum Computing and Department of Physics and Astronomy, University of Waterloo, Waterloo, ON N2L 3G1, Canada}
\author{Carlton M.~Caves}
\affiliation{Center for Quantum Information and Control, University of New Mexico, MSC07-4220, Albuquerque, New Mexico 87131-0001, USA}
\affiliation{Centre for Engineered Quantum Systems, School of Mathematics
and Physics, The University of Queensland, St.~Lucia, Brisbane 4072,
Australia}

\begin{abstract}
Any evolution described by a completely positive trace-preserving linear map can be imagined as arising from the interaction of the evolving system with an initially uncorrelated ancilla. The interaction is given by a joint unitary operator, acting on the system and the ancilla. Here we study the properties such a unitary operator must have in order to force the choice of a physical---that is, positive---state for the ancilla if the end result is to be a physical---that is, completely positive---evolution of the system.
\end{abstract}

\maketitle

\section{Introduction}
\label{sec:introduction}

Consider a primary system $A$, with Hilbert space $\sH_A$ of dimension $d_A$, which is subjected to a superoperator $\sE$, and let $B$ be an ancillary system, with Hilbert space $\sH_B$ of dimension $d_B$.  It is a commonplace that if the action of $\sE$ can be written as
\begin{equation}\label{eq:ancillamodel}
\sE(\rho_A)=\tr_B\n \big( U \rho_A\otimes \sigma_B U^{\dag}\big)\;,
\end{equation}
where $U$ is a joint unitary operator acting on $A$ and $B$ and $\sigma_B$ is a physical state of $B$ (a unit-trace, positive operator, i.e., a density operator), then $\sE$ is a trace-preserving, completely positive (CP) map and is called a \emph{trace-preserving quantum operation}~\cite{Nielsen2000a, Choi1975a}.  Moreover, any quantum operation can be written in the form~(\ref{eq:ancillamodel}) for some joint unitary $U$ and some ancilla state $\sigma_B$.  Quantum operations are thus the superoperator maps that can be realized physically~\cite{Buscemi2003a}.  The form~(\ref{eq:ancillamodel}) for a quantum operation is sometimes called a \emph{Stinespring extension}~\cite{Stinespring1955a}, and we call it here an \emph{ancilla model}.

In this paper we are interested in a different question.  Suppose we know that $\sE$, written as in Eq.~(\ref{eq:ancillamodel}) with $U$ a unitary operator, is a trace-preserving quantum operation.  Can we conclude that $\sigma_B$ is a density operator?  Generally not, as the case of a unitary transformation $U_A$ of $A$ makes immediately clear.  Then we have $U=U_A\otimes I_B$ and $\sE(\rho_A)=\tr_B(\sigma_B)U_A\rho_A U_A^\dagger$, so $\sigma_B$ can be any unit-trace operator.  A natural question asks for the conditions on $U$ such that $\sE$ being a trace-preserving quantum operation implies that $\sigma_B$ is a density operator.  Here we formulate and prove a theorem that answers this question.

In Sec.~\ref{sec:math}, we spell out the problem clearly and introduce mathematical concepts that are useful in addressing the problem and proving our main result.  Section~\ref{sec:theorem} states and proves our theorem, and Sec.~\ref{sec:examples} explores a number of examples that illustrate features of the main result.  Section~\ref{sec:tomography} demonstrates a curious connection to the notion of indirect state tomography, i.e., tomography on the system that can determine the state of the ancilla, and Sec.~\ref{sec:conclusion} concludes.

\section{Mathematical background}
\label{sec:math}

Before stating our theorem, we introduce some mathematical notation and concepts.  Consider a system $Q$ with Hilbert space $\sH_Q$ of dimension $d_Q$.  We will consistently refer to arbitrary elements of $\sH_Q$ as vectors and to vectors normalized to unity as state vectors.  We refer to density operators, i.e., unit-trace, positive operators, as states; pure states are rank-one density operators, i.e., one-dimensional projectors corresponding to state vectors.  We sometimes use a bra-ket notation for operators on $\sH_Q$, i.e., $O=|O)$ and $O^\dagger=(O|$, and we denote the Hilbert-Schmidt operator inner product by $(N|O)=\tr(N^\dagger O)$.  We use a tilde to distinguish operators that are normalized to unity, i.e., $(\tilde O|\tilde O)=\tr(O^\dagger O)=1$.

A superoperator is a linear map on operators.  Any superoperator $\sF_Q$ can be written as
\begin{equation}
\sF_Q=\sum_{\alpha,\beta}\sF_{\alpha\beta}\tilde\tau_\alpha\odot\tilde\tau_\beta^\dagger
=\sum_{\alpha,\beta}\sF_{\alpha\beta}|\tilde\tau_\alpha)(\tilde\tau_\beta|\;,
\end{equation}
where the operators $\tilde\tau_\alpha$ make up an orthonormal operator basis.  The \emph{ordinary action\/} of $\sF_Q$ on an operator $O$ corresponds to replacing the $\odot$ by $O$:
\begin{equation}
\sF_Q(O)=\sum_{\alpha,\beta}\sF_{\alpha\beta}\tilde\tau_\alpha O\tilde\tau_\beta^\dagger\;.
\end{equation}
The \emph{left-right action\/} corresponds to using the bra-ket form of $\sF_Q$ to operate to the left or right on operators.  A superoperator $\sF_Q$ has a \emph{Kraus decomposition\/}~\cite{Kraus1983a} if it can be written as
\begin{equation}
\sF_Q=\sum_\alpha K_\alpha\odot K_\alpha^\dagger=\sum_\alpha|K_\alpha)(K_\alpha|\;.
\end{equation}
The operators $K_\alpha$ are called \emph{Kraus operators}.  Having a Kraus decomposition is equivalent to $\sF_Q$ being positive in its left-right action.

The standard way of defining complete positivity of a superoperator invokes a reference system $R$.  A superoperator $\sF_Q$ is \emph{completely positive\/} if $\sI_R\otimes\sF_Q$ maps positive operators to positive operators for all reference systems~$R$, where $\sI_R=I_R\odot I_R$ is the identity superoperator on $R$.  It is clear that if $\sF_Q$ is left-right positive, i.e., has a Kraus decomposition, $\sI_R\otimes\sF_Q$ maps positive operators to positive operators, thus making $\sF_Q$ completely positive.  Moreover, if $R$ is chosen to have the same dimension as $Q$ and if we introduce an (unnormalized) maximally entangled vector,
\begin{equation}
|\Psi\rangle=\sum_{n=1}^{d_Q}|g_n\rangle\otimes|e_n\rangle\;,
\end{equation}
where $\{|g_n\rangle\}$ and $\{|e_n\rangle\}$ are orthonormal bases on $R$ and $Q$, then we can define one-to-one, onto maps of operators on $Q$ to vectors on $RQ$ and of superoperators on $Q$ to operators on $RQ$ via
\begin{align}
|O_{RQ}\rangle\equiv I_R\otimes O_Q|\Psi\rangle
\quad&\Longleftrightarrow\quad
\langle g_n,e_m|O_{RQ}\rangle=\langle e_m|O_Q|e_n\rangle=
\bigl(|e_m\rangle\langle e_n|\bigl|\bigr.O_Q\bigr)\;,\\
F_{RQ}\equiv \sI_R\otimes\sF_Q\bigl(|\Psi\rangle\langle\Psi|\bigr)
\quad&\Longleftrightarrow\quad
\langle g_n,e_m|F_{RQ}|g_{n'},e_{m'}\rangle=
\bigl\langle e_m\bigl|\sF_Q\bigl(|e_n\rangle\langle e_{n'}|\bigr)\bigr| e_{m'}\bigr\rangle=
\bigl(|e_m\rangle\langle e_n|\bigl|\sF_Q\bigr||e_{m'}\rangle\langle e_{n'}|\bigr)\;.
\label{eq:Jam}
\end{align}
The isomorphism between superoperators on $Q$ and operators on $RQ$ is called the \emph{Choi-Jamio{\l}kowski isomorphism}~\cite{Choi1975a,Jamiokowski1972a}.  The second form of Eq.~(\ref{eq:Jam}) shows that the left-right positivity of $\sF_Q$ is equivalent to the positivity of $F_{RQ}$.  Hence, we have the standard result that $\sF_Q$ is completely positive if and only if it is left-right positive.

Given any orthonormal basis of operators, $\{\tilde\tau_\alpha\}$, acting on $\sH_Q$, we can define the unit superoperator acting in the left-right sense:
\begin{equation}
\bI_Q\equiv \sum_\alpha\tilde\tau_\alpha\odot\tilde\tau_\alpha^\dagger=\sum_\alpha|\tilde\tau_\alpha)(\tilde\tau_\alpha|\;.
\end{equation}
By considering an outer-product basis, $\{\,\tilde\tau_{jk}=|e_j\rangle\langle e_k|\,\}$ for some orthonormal basis, $\{\,|e_j\rangle\,\}$, it is easy to check that
\begin{equation}
\bI_Q(O)=\sum_{j,k}\tilde\tau_{jk} O\tilde\tau_{jk}^\dagger=\tr(O)\,I_Q\;.
\end{equation}

The setting of our theorem is that $U$ is a joint unitary operator on $A$ and $B$ and $\sigma_B$ is an operator on $B$.  We define a superoperator on $A$ by its action~(\ref{eq:ancillamodel}) on density operators.  We want $\sE$ to map density operators to density operators, so it must at least be Hermiticity- and trace-preserving.  Requiring that $\sE(\rho_A)$ be Hermitian does not imply that $\sigma_B$ is Hermitian, but it does imply that $\sE(\rho_A)=\tr_{B}\n \big( U \rho_A\otimes\frac{1}{2}(\sigma_B+\sigma_B^\dagger) U^{\dag}\big)$.  Since the Hermitian ancilla operator $\frac{1}{2}(\sigma_B+\sigma_B^\dagger)$ gives the same map as $\sigma_B$, we can assume that $\sigma_B$ is Hermitian without loss of generality.  Requiring that $\sE$ be trace preserving implies that $\sigma_B$ has unit trace.

We will need the Schmidt decomposition of $U$~\cite{Nielsenthesis},
\begin{equation}\label{eq:Schmidtdecomposition}
U=\sum_{n=1}^{R_U}A_n\otimes B_n\;,
\end{equation}
where $R_U$ is the Schmidt rank of $U$.  The Schmidt operators $A_n$
are nonzero and orthogonal,
\begin{equation}
(A_n|A_m)=\tr_A(A_n^\dagger A_m)=(A_n|A_n)\delta_{nm}\;,
\end{equation}
and likewise for the Schmidt operators $B_n$.  Sometimes we want to deal with normalized operators,
\begin{equation}\label{eq:orthonormalSchmidt}
\tilde{A}_n\equiv\frac{A_n}{\sqrt{(A_n|A_n)}}
\qquad\mbox{and}\qquad
\tilde{B}_n\equiv\frac{B_n}{\sqrt{(B_n|B_n)}}\;.
\end{equation}
We can normalize either set, $\{A_n\}$ or $\{B_n\}$, by absorbing the normalization constant into the other set.  That $U$ is unitary means that
\begin{equation}\label{eq:unitarity}
I=U^\dagger U=\sum_{n,m=1}^{R_U}A_n^\dagger A_m\otimes B_n^\dagger B_m\;,
\end{equation}
which, by taking the partial trace on $A$, implies that
\begin{equation}\label{eq:unitaritytwo}
I_B=\frac{1}{d_A}\sum_{n=1}^{R_U}\,(A_n|A_n) B_n^\dagger B_n\;.
\end{equation}
Taking the trace on $B$ then gives
\begin{equation}\label{eq:tracetrace}
d_Ad_B=\sum_{n=1}^{R_U}\,(A_n|A_n)(B_n|B_n)\;.
\end{equation}

We will need the span of the Schmidt operators $B_n$,
\begin{equation}\label{eq:spanB}
\sB\equiv\mbox{Span}\bigl\{\,B_n\,\bigm|\,n=1, 2,\cdots,R_U\,\bigr\}\;,
\end{equation}
and the operator orthocomplement of $\sB$,
\begin{equation}\label{eq:oportho}
\sO_B\equiv\bigl\{\,O_B\,\bigm|\,\tr_B(B^\dagger O_B)=0\;\mbox{$\forall$ $B\in \sB$}\,\bigr\}
=\bigl\{\,O_B\,\bigm|\,\tr_B(U^\dagger O_B)=0\,\bigr\}.
\end{equation}
Notice that $R_U+\dim(\sO_B)=d_B^{\,2}$.

It will also be useful to introduce a set of positive operators associated with $\sB$,
\begin{equation}
\sC\equiv\bigl\{\, B^\dagger B\,\bigm|\, B\in \sB\,\bigr \}\;.
\end{equation}
The set $\sC$ is not a subspace.  It is a cone---i.e., any nonnegative multiple of an element of $\sC$ is also an element---which is a subset of the cone of positive operators on $B$.  It is clearly a closed set.  We let $\sC_1$ denote the set of unit-trace elements in $\sC$; $\sC_1$ is a closed and bounded subset of the closed and bounded set of density operators on $B$.  Neither $\sC$ nor $\sC_1$ is necessarily convex.

For the remainder of this section, we deal only with system $B$, so we generally omit the subscript $B$ on vectors, operators, and superoperators. The most important set in our considerations is the set of vectors $|\phi\rangle\in\sH_B$ such that $|\phi\rangle\langle\phi|\notin \sC$.  Any nonzero multiple of such a $\ket{\phi}$ is also in this set, but we emphasize that the set is not a subspace.  For this reason, it is most convenient to represent the vectors in this set by their normalized counterparts (state vectors), so we define the following set:
\begin{equation}\label{eq:SBc}
S_B\equiv
\bigl\{\,\mbox{state vectors $|\phi\rangle\in\sH_B$}\,\bigm|\,\ket{\phi}\bra{\phi}\notin\sC\,\bigr\}\;.
\end{equation}
The set $S_B$ is a property of the Schmidt operators $B_n$ and, hence, a property of $U$.

Since we have restricted $S_B$ to normalized vectors, we can replace the definition by
\begin{equation}\label{eq:SBbest}
S_B=\bigl\{\,\mbox{state vectors $|\phi\rangle\in\sH_B$}\,\bigm|\,\ket{\phi}\bra{\phi}\notin\sC_1\,\bigr\}\;.
\end{equation}
We can think of $S_B$ as the set of pure states not in~$\sC_1$.  Because $\sC_1$ is a closed set, we can say a bit more.  If the one-dimensional projector $\ket{\phi}\bra{\phi}$ is not in $\sC_1$, then it lies some minimal distance away from all elements of $\sC_1$, which means that there exists a $\delta>0$ such that $\bra{\phi}C\ket{\phi}\le1-\delta$ for all $C\in\sC_1$.  Slightly more generally, we can say that
\begin{equation}\label{eq:projcondition}
\mbox{state vector $\ket{\phi}\in\sH_B$ such that $\ket{\phi}\bra{\phi}\notin\sC_1$}
\;\Longrightarrow\;
\mbox{$\exists$ a $\delta>0$ such that $\bra{\phi}C\ket{\phi}\le\tr(C)(1-\delta)$ $\;\forall$ $C\in\sC$.}
\end{equation}

It will turn out to be useful to have available equivalent definitions of $S_B$.  Suppose $\ket{\psi}\bra{\phi}=B\in\sB$; then $\langle\psi\ket{\psi}\ket{\phi}\bra{\phi}=B^\dagger B=C$, which if $\ket{\psi}$ is not the zero vector, implies that $\ket{\phi}\bra{\phi}\in\sC$.  Moreover, if $\ket{\phi}\bra{\phi}=C=B^\dagger B$ for some $C\in\sC$ (and $B\in\sB$), the polar-decomposition theorem asserts that $B$ is rank one, and thus $B=\ket{\psi}\bra{\phi}$ for some normalized $\ket{\psi}\in\sH_B$.  Thus we have the following equivalence,
\begin{equation}
\mbox{$\ket{\psi}\bra{\phi}\in\sB$ for some nonzero $\ket{\psi}\in\sH_B$}
\;\Longleftrightarrow\;
\ket{\phi}\bra{\phi}\in\sC\;,
\end{equation}
and its contrapositive,
\begin{equation}\label{mainequiv2}
\mbox{$\forall$ nonzero $\ket{\psi}\in\sH_B$, $\ket{\psi}\bra{\phi}\notin\sB$}
\;\Longleftrightarrow\;
\ket{\phi}\bra{\phi}\notin\sC\;.
\end{equation}

Consider now the subspace of $\sH_B$ generated by the operators in the orthocomplement $\sO_B$ acting on a particular vector $|\phi\rangle\in\sH_B$:
\begin{equation}\label{eq:SphiB}
\sO_B\ket{\phi}\equiv
\bigl\{\,O|\phi\rangle\,\bigm|\,O\in\sO_B\,\bigr\}\;.
\end{equation}
It is easy to see that $\sO_B\ket{\phi}$ is a subspace; it is the span of vectors $O_n|\phi\rangle$, where the set $\{O_n\}$ consists of any operators that span $\sO_B$.  Thus we clearly have $\dim(\sO_B\ket{\phi})\le\dim(\sO_B)$.  We denote the orthocomplement of
$\sO_B\ket{\phi}$ by $(\sO_B\ket{\phi})_\perp$.

Because $\sB$ and $\sO_B$ are orthogonal subspaces, we have
\begin{equation}
\ket{\psi}\bra{\phi}\in\sB
\;\Longleftrightarrow\;
\mbox{$\forall$ $O\in\sO_B$, $0=\tr\!\big(O^\dagger  \ket{\psi}\bra{\phi}\big)
=\bra{\phi} O^\dagger\ket{\psi}$}
\;\Longleftrightarrow\;
\ket{\psi}\in(\sO_B\ket{\phi})_\perp
\end{equation}
and the contrapositive,
\begin{equation}
\ket{\psi}\bra{\phi}\notin\sB
\;\Longleftrightarrow\;
\ket{\psi}\notin(\sO_B\ket{\phi})_\perp\;.
\end{equation}
This gives us the following equivalence:
\begin{equation}\label{eq:mainequiv}
\mbox{$\forall$ nonzero $\ket{\psi}\in\sH_B$, $\ket{\psi}\bra{\phi}\notin\sB$}
\;\Longleftrightarrow\;
\mbox{$\sO_B\ket{\phi}=\sH_B$}\;.
\end{equation}

Thus 
 because of Eqs.~(\ref{mainequiv2}) and~(\ref{eq:mainequiv}), the following two ways of defining $S_B$ are equivalent to Eqs.~(\ref{eq:SBc}) and (\ref{eq:SBbest}):
\begin{align}\label{eq:SB}
S_B
&=\bigl\{\,\mbox{state vectors $|\phi\rangle\in\sH_B$}\,\bigm|\,\mbox{$\forall$ nonzero $\ket{\psi}\in \sH_B$, $\ket{\psi}\bra{\phi}\notin\sB$}\,\bigr\}\nonumber\\
&=\bigl\{\,\mbox{state vectors $|\phi\rangle\in\sH_B$}\,\bigm|\,\sO_B\ket{\phi}=\sH_B\,\bigr\}\;.
\end{align}

It is useful to characterize $S_B$ in yet another way.  For this purpose, consider the orthonormal Schmidt operators, $\{\tilde B_n$, $n=1,\ldots,R_U\,\}$, of~\req{eq:orthonormalSchmidt}.  Complete a basis of orthonormal operators by adding a set of orthonormal operators, $\{\,\tilde O_n\mid n=R_U+1,\ldots,d_B^{\,2}\,\}$, from the orthocomplement $\sO_B$.  We have
\begin{equation}
\bI=\sum_{n=1}^{R_U}\tilde B_n\odot\tilde B_n^\dagger
+\sum_{n=R_U+1}^{d_B^{\,2}}\tilde O_n\odot\tilde O_n^\dagger\;,
\end{equation}
and for state vectors $|\phi\rangle\in\sH_B$,
\begin{equation}
\label{eq:normcond}
I=\bI\bigl(|\phi\rangle\langle\phi|\bigr)
=\sum_{n=1}^{R_U}\tilde B_n|\phi\rangle\langle\phi|\tilde B_n^\dagger
+\sum_{n=R_U+1}^{d_B^{\,2}}\tilde O_n|\phi\rangle\langle\phi|\tilde O_n^\dagger\;.
\end{equation}
A state vector $|\phi\rangle$ is in $S_B$ if and only if the second sum on the rightmost side of Eq.~(\ref{eq:normcond}) is a full-rank operator, i.e., has no zero eigenvalues.  This happens if and only if the first sum on the rightmost side of Eq.~(\ref{eq:normcond}) is strictly less than $I$; i.e., it has largest eigenvalue strictly less than unity.  Thus we can characterize $S_B$ in the following way:
\begin{equation}
\label{eq:S_Bopnorm}
S_B=
\biggl\{\,\mbox{state vectors $|\phi\rangle\in\sH_B$}\biggm|\,1>
\biggl\|\,\sum_{n=1}^{R_U}\tilde B_n|\phi\rangle\langle\phi|\tilde B_n^\dagger\,\biggr\|_\infty\,\biggr\}\;.
\end{equation}
Here $\|M\|_\infty$ is the operator norm that denotes the largest
eigenvalue of $\sqrt{M^\dagger M}$, i.e., the largest singular value of $M$ (for positive operators, as here, the largest eigenvalue of $M$).

\section{Main theorem}
\label{sec:theorem}

The content of our theorem is that any rank-one projector not in $\sC_1$ can be used to construct an unphysical $\sigma_B$ that gives a completely positive map $\sE$.  Formally, we can state the theorem in the following way.
\vspace{6pt}

\noindent
{\bf Main theorem.}
Let $U$ be a joint unitary operator on $A$ and $B$ and $\sigma_B$ be a unit-trace Hermitian operator on $B$.  Define a Hermicity- and trace-preserving superoperator $\sE$ on $A$ by $\sE(\rho_A)=\tr_{B}\n\big( U \rho_A\otimes \sigma_B U^{\dag}\big)$.  $\sC_1$ contains all rank-one projectors, i.e., $S_B$ is the empty set, if and only if requiring that $\sE$ be a quantum operation implies that $\sigma_B$ is a density operator. In symbols,
\begin{align}
\mbox{$\forall$ state vector $|\phi_B\rangle\in\sH_B$},\;\ket{\phi_B}\bra{\phi_B}\in\sC_1 &\;\Longleftrightarrow\; S_B=\emptyset\nonumber\\
&\;\Longleftrightarrow\;
\Bigl(\mbox{$\sE$ a quantum operation $\Longrightarrow$ $\sigma_B$ is a density operator}\Bigr)\;.
\end{align}
We use P to denote the property of $U$ that $S_B$ is the empty set. $\neg$P is thus that $S_B$ is not the empty set, i.e., that there are pure states not in $\sC_1$.  Any $|\phi_B\rangle\in S_B$ can be used as an eigenstate of $\sigma_B$ with negative eigenvalue in an ancilla model that gives a completely positive $\sE$.
\vspace{6pt}

The following properties imply P and thus are sufficient to ensure that $\sigma_B$ must be a density operator:
\begin{itemize}
\begin{item}[p1.]
$\Bigl(\,\tr_B(U^\dagger O_B)=0$ $\Longrightarrow$ $O_B=0\,\Bigr)$, which is equivalent to
$\dim(\sO_B)=0$ and
$R_U=d_B^{\,2}$.
\end{item}
\begin{item}[p2.]
$R_U>d_B^{\,2}-d_B$, which is equivalent to
$\dim(\sO_B)<d_B$.
This property implies P because
$\dim(\sO_B\ket{\phi_B})\le\dim(\sO_B)$.
\end{item}
\end{itemize}
\noindent
The following properties imply $\neg$P and thus are sufficient to ensure that there is a nonpositive $\sigma_B$ that leads to a completely positive $\sE$:
\begin{itemize}
\begin{item}[q1.]
$\exists$ a non-vanishing Hermitian operator $\Sigma_B$ such that $0=\tr_B(B_n\Sigma_B B_m^\dagger)=\tr_B(B_m^\dagger B_n\Sigma_B)$ $\forall$ $n,m$.  This is equivalent to $\tr_B(B\Sigma_B B^\dagger)=0$ for all $B\in\sB$ and thus to $\tr_B(C\Sigma_B)=0$ for all $C\in\sC$.  If a non-Hermitian operator satisfies this property, so does its conjugate, thus allowing us to construct a Hermitian operator that satisfies the property.  That $\Sigma_B$ is necessarily traceless follows from \req{eq:unitaritytwo}.  Suppose we add $\mu\Sigma_B$, where $\mu$ is any real number, to a density operator $\sigma_B$ that is diagonal in the same basis as $\Sigma_B$, i.e., $\sigma'_B=\sigma_B+\mu\Sigma_B$.  Then $\sigma'_B$ has unit trace, and the matrix $G$ of~\req{eq:G}---and, hence, the quantum operation $\sE$---remains the same, but by making $|\mu|$ large enough, we can give $\sigma'_B$ negative eigenvalues.  All the eigenvectors of $\Sigma_B$ are elements of~$S_B$.
\end{item}
\begin{item}[q2.]
$\dim\Bigl(\mbox{Span}\bigl\{\,B_m^\dagger B_n\,\bigm|\,m,n=1,\ldots,R_U\,\bigr\}\Bigr)<d_B^{\,2}$.  This is equivalent to q1.
\end{item}
\begin{item}[q3.]
$R_U<d_B$, which is equivalent to $\dim(\sO_B)\ge d_B^{\,2}-d_B$.  This property implies q2, since $\dim\Bigl(\mbox{Span}\bigl\{\,B_m^\dagger B_n\,\bigr\}\Bigr)\le R_U^2$.
\end{item}
\end{itemize}
\noindent
Now we are ready to give the proof.

\vspace{6pt}
{\bf Proof.} We first reformulate the condition for complete positivity of $\sE$ in terms of the Schmidt operators on $B$.  Plugging \req{eq:Schmidtdecomposition} into \req{eq:ancillamodel}, we find
\begin{equation}\label{eq:ESchmidt}
\sE=\sum_{n,m=1}^{R_U}G_{nm}A_n\odot A_m^\dagger
=\sum_{n,m=1}^{R_U}G_{nm}|A_n)(A_m|\;,
\end{equation}
where
\begin{equation}\label{eq:G}
G_{nm}\equiv(A_n|\sE|A_m)=\tr_B(B_n\sigma_B B_m^\dagger)\;.
\end{equation}
Complete positivity of $\sE$ is the statement that $\sE$ is positive in its left-right action.  This is equivalent to the positivity of the matrix $G$, which is that $\sum_{n,m}c_nG_{nm}c_m^*\ge0$ for all coefficients $c_n$, and this becomes the property
\begin{equation}
\mbox{$\tr_B(B\sigma_B B^\dagger)\ge0$ $\;\forall$ $B\in\sB$.}
\end{equation}
We have reduced our considerations to quantities defined on system~$B$, so we often drop the subscript $B$ for the rest of the proof.  The complete positivity property of $\sE$ can now be written in the compact form
\begin{equation}\label{eq:cpcondition3}
\tr( C \sigma ) \geq 0 \mbox{~$\;\forall$ $C\in\sC$}\;.
\end{equation}

If P holds, $\sC$ contains all one-dimensional projectors, so the complete positivity condition~(\ref{eq:cpcondition3}) immediately implies that $\sigma\ge0$.  We prove the converse in the contrapositive; i.e., we show that if $\neg$P is true, we can construct a nonpositive (unphysical) $\sigma$ that is consistent with~\req{eq:cpcondition3}.  Since P is not true, there is a state vector $\ket{\phi}$ such that $\ket{\phi}\bra{\phi}\notin\sC$ and, hence, by~\req{eq:projcondition}, a $\delta>0$ such that $\bra{\phi}C\ket{\phi}\le\tr(C)(1-\delta)$ for all $C\in\sC$.  We lose nothing by also requiring that $\delta<1/2$.  For any such state vector, we can construct a nonpositive $\sigma$ that satisfies~\req{eq:cpcondition3} by letting $\ket{\phi}$ be an eigenvector with a negative eigenvalue $-\epsilon$:
\begin{equation}
\sigma=-\epsilon\, \ket{\phi}\bra{\phi}+\frac{1+\epsilon}{d_B-1}\,
\big( I-\ket{\phi}\bra{\phi}\big)\;,\quad\epsilon>0.
\end{equation}
For all $C\in\sC$, we have
\begin{equation}
\tr(C\sigma)=-\frac{1+\epsilon d_B}{d_B-1}\,\bra{\phi}C\ket{\phi}
+\frac{1+\epsilon}{d_B-1}\,\tr(C)
\ge\left(\delta\frac{1+\epsilon d_B}{d_B-1}-\epsilon\right)\tr(C)\;.
\end{equation}
By choosing $\epsilon\leq\delta/(d_B-1-\delta d_B)$, we can ensure that the complete positivity condition~(\ref{eq:cpcondition3}) is satisfied for this nonpositive $\sigma$, thereby completing the proof. $\blacksquare$

\section{Examples}
\label{sec:examples}

\emph{Example~1.} $U=U_A\otimes U_B$.  In this case, any unit-trace operator $\sigma_B$ gives the same (unitary) quantum operation.  There is a single Schmidt operator, $U_B$, on $B$ ($R_U=1$).  The subspace $\sB$ is the one-dimensional subspace of all multiples of $U_B$, $\sC$ is the cone of nonnegative multiples of $I_B$, $\sC_1=\{\,I_B/d_B\,\}$, $\sO_B$ is the ($d_B^{\,2}-1$)-dimensional space of operators of the form $U_B O$, where $O$ is any traceless operator, and $S_B$ contains all state vectors.
\vspace{6pt}

\noindent
\emph{Example~2.}  The SWAP operator for two systems of the same Hilbert-space dimension:
\begin{equation}\label{eq:SWAP}
U=\mbox{SWAP}=\sum_{j,k}|e_k,f_j\rangle\otimes\langle e_j,f_k|
=\sum_{j,k}|e_k\rangle\langle e_j|\otimes|f_j\rangle\langle f_k|\;.
\end{equation}
Here the vectors $|e_j\rangle$ and $|f_k\rangle$ make up any orthonormal bases for $A$ and $B$.  The SWAP operator acts on product states as $U\rho_A\otimes\sigma_B U^\dagger=\sigma_A\otimes\rho_B$, so it is clear that $\sigma_B$ must be a density operator.  Equation~(\ref{eq:SWAP}) is a Schmidt decomposition of \hbox{SWAP}.  Schmidt operators on $B$ can be taken to be the outer products $|f_j\rangle\langle f_k|$.  These operators span the operator space on $B$, so $\sB$ contains all operators, $\sC$ is the cone of positive operators, $\sC_1$ is the set of density operators, $\sO_B$ consists only of the zero operator, and $S_B$ is empty.
\vspace{6pt}

\noindent
\emph{Example~3.} Two qubits with a joint unitary
\begin{equation}\label{eq:qubitexample}
U=e^{iX\otimes X\theta/2}=I\otimes I\cos(\theta/2) + iX \otimes X\sin(\theta/2)\;,
\quad 0<\theta<\pi\;.
\end{equation}
This joint unitary has Schmidt rank~2.  For the unitary of \req{eq:qubitexample}, the two Schmidt operators for $B$ can be taken to be $I$ and $X$, $\sO_B={\rm Span}\{Y,Z\}$, $\sC$ consists of all operators of the form $r(I+sX)$, where $r\ge0$ and $|s|\le1$, $\sC_1$ is the subset of $\sC$ with $r=1/2$ (i.e., the density operators whose eigenstates are the eigenstates of $X$), and $S_B$ contains all state vectors except the eigenstates of $X$.

If we write $\sigma_B$ as $\sigma_B=\frac{1}{2}(I+\boldsymbol{s\cdot\sigma})=\frac{1}{2}(I+s_xX_B+s_yY_B+s_zZ_B)$, where $\boldsymbol{\sigma}$ is the vector of Pauli operators on~$B$, it is straightforward to show that the superoperator~(\ref{eq:ancillamodel}) on~$A$ is
\begin{align}
\sE=\cos^2\!(\theta/2)I\odot I+i s_x\sin(\theta/2)\cos(\theta/2)(X\odot I-I\odot X)+\sin^2\!(\theta/2)X\odot X=p_+V\odot V^\dagger+p_- XV\odot V^\dagger X\;,
\end{align}
where $p_{\pm}=\frac{1}{2}(1\pm T)$, $T=\sqrt{\cos^2\!\theta+s_x^2\sin^2\!\theta}$, $V=e^{iX\alpha/2}$, and $\tan\alpha=s_x\tan\theta$.  This map is a convex combination of two orthogonal unitaries, $V$ and $XV$, and is completely positive if and only if $T\le1$, i.e., $|s_x|\le1$.  The point here is that $s_y$ and $s_z$ can be anything without changing $\sE$.  Complete positivity only constrains $s_x$, in agreement with our conclusion that the only state vectors not in $S_B$ are the eigenstates of $X_B$.
\vspace{6pt}

Any two-qubit joint unitary has Schmidt rank~1, 2, or 4~\cite{Nielsen2003a}.  A Schmidt-rank-1 unitary is a product unitary and thus is covered by Example~1.  A Schmidt-rank-2 unitary is equivalent under local unitaries to \req{eq:qubitexample} and thus covered by Example~3.  A Schmidt-rank-4 unitary has full Schmidt rank; $C_1$ is the set of all density operators, and $S_B$ is the empty set.  Since $\sC_1$ is convex in all three cases, we can conclude that for two-qubit unitaries, $\sC$ and $\sC_1$ are always convex sets.
\vspace{6pt}

\noindent
\emph{Example~4.} A qutrit and a qubit with joint unitary
\begin{equation}\label{eq:tritbit}
U= \proj{0} \otimes I + \proj{1} \otimes X + \proj{2} \otimes Z\;.
\end{equation}
This joint unitary has Schmidt rank~3.  The Schmidt operators on $B$ can be taken to be $I$, $X$, and $Z$, $\sO_B$ consists of all multiples of $Y$, $\sC$ is the cone of positive operators, $\sC_1$ is the set of all density operators, and $S_B$ is empty.  That $U$ satisfies p2 makes clear that $S_B$ is empty.
\vspace{6pt}

For the remaining examples, we refer to an orthonormal basis, $\{\,\ket{j},\;j=0,\ldots d_B-1\,\}$, on system~$B$, we denote the rank-one outer products by $\tau_{jk}=\ket{j}\bra{k}$, and we use the following operators on~$B$, defined for $j<k$:
\begin{align}
I_{jk}&\equiv\tau_{jj}+\tau_{kk}\;,\\
Z_{jk}&\equiv\tau_{jj}-\tau_{kk}\;,\\
X_{jk}&\equiv\tau_{jk}+\tau_{kj}\;,\\
Y_{jk}&\equiv-i(\tau_{jk}-\tau_{kj})\;.
\end{align}
These are the Pauli operators for the two-dimensional subspace spanned by $\ket{j}$ and $\ket{k}$.  The following operators make up a set of $d_B^{\,2}$ linearly independent operators, which span the space of operators on $B$: the unit operator $I$, and the operators $Z_{0j}$, $j=1,\ldots,d_B-1$, span the space of diagonal operators; the operators $X_{jk}$, numbering $d_B(d_B-1)/2$, and the operators $Y_{jk}$, numbering $d_B(d_B-1)/2$, span the space of off-diagonal operators.   These operators are pairwise orthogonal, except that $\tr(Z_{0j}Z_{0k})=1+\delta_{jk}$.

We can introduce an associated set of $d_B^{\,2}-1$ unitary operators:
\begin{align}
U_j&\equiv I-I_{0j}+Z_{0j}\;,\quad j=1,\ldots,d_B-1,\\
V_{jk}&\equiv I-I_{jk}+X_{jk}\;,\quad j<k,\\
W_{jk}&\equiv I-I_{jk}+Y_{jk}\;,\quad j<k.
\end{align}
The unitary $U_j$ is diagonal, with ones everywhere on the diagonal, except in the $j$th position, which has a $-1$.  The operators $V_{jk}$ and $W_{jk}$ act like the Pauli $X$ and $Y$ in the
two-dimensional subspace spanned by $\ket j$ and $\ket k$, and act like the identity operator on the orthocomplement of this subspace.  Together with the identity operator $I$, these operators are linearly independent and span the space of operators on $B$, but they are not orthogonal.  To see that they span the space, one notes that the diagonal operators satisfy
\begin{align}
\tau_{jj}&=\frac{1}{2}(I-U_j)\;,\quad j=1,\ldots,d_B-1,\\
\tau_{00}&=\frac{1}{d_B-1}\sum_{j=1}^{d_B-1}U_j-\frac{d_B-3}{d_B-1}\sum_{j=1}^{d_B-1}\tau_{jj}\;.
\end{align}
These expressions allow us to write any diagonal operator in terms of $I$ and the unitaries $U_j$.  Writing the diagonal operators $I-I_{jk}$ in terms of $I$ and the unitaries $U_j$, we find that the operators $X_{jk}$ and $Y_{jk}$, which span the off-diagonal subspace, can be expanded in terms of our set of unitaries.
\vspace{3pt}

\noindent
\emph{Example~5.} This example shows that P can be satisfied even though $R_U=3\m d_B-2$ is linear in $d_B$.  The idea is to introduce a unitary $U$ such that the set $\cal B$ contains the operator subspace spanned by the operators $\tau_{0j}$ for $j=0,\ldots,d_B-1$, i.e., contains all operators of the form
\begin{equation}
B=\lambda\sum_{j=0}^{d_B-1}b_j^*\tau_{0j}\;,
\end{equation}
where $\lambda\ge0$ and the expansion coefficients $b_j$ are normalized to unity.  This means that $\cal C$ contains all operators of the form
\begin{equation}
B^\dagger
B=\lambda^2\sum_{j,k}b_jb_k^*\tau_{jk}=\lambda^2\ket{\phi}\bra{\phi}\;
,
\end{equation}
where
\begin{equation}\label{eq:bvector}
\ket{\phi}=\sum_jb_j\ket{j}
\end{equation}
is a state vector in ${\cal H}_B$.  Thus ${\cal C}_1$ contains all pure states, and $S_B=\emptyset$.

To do this, we assume that system~$A$ has dimension $3d_B-2$, with an orthonormal basis of vectors $\ket0$, $\ket{r_j}$, $\ket{s_j}$, and $\ket{t_j}$, where $j=1,\ldots,d_B-1$, and we introduce the following controlled unitary on $A$ and $B$:
\begin{equation}
U=\ket0\bra0\otimes I+\sum_{j=1}^{d_B-1}\Bigl(\ket{r_j}\bra{r_j}\otimes
U_j+\ket{s_j}\bra{s_j}\otimes V_{0j}
+\ket{t_j}\bra{t_j}\otimes W_{0j}\Bigr)\;.
\end{equation}
Because the projectors corresponding to the orthonormal basis on $A$ are orthogonal and, hence, linearly independent, ${\cal B}$ is the span of the target
unitaries, and the operator subspace ${\cal O}_B$ consists of all operators that are orthogonal to all of the target unitaries.  By the same argument as above, we then have
\begin{equation}\label{eq:B5}
{\cal B}=\mbox{Span}\bigl\{\,
\mbox{$\tau_{00}$; $\tau_{jj}$, $\tau_{0j}$, and $\tau_{j0}$, for
$j=1,\ldots,d_B-1$}\,\bigr\}\;.
\end{equation}
As noted above, in this situation, ${\cal C}_1$ contains all pure states and thus P is satisfied.

We can go a bit further and characterize the entire set ${\cal C}_1$.  An arbitrary element of $\cal B$ has the form
\begin{equation}
B=\lambda\sum_{j=0}^{d_B-1}b_j^*\tau_{0j}
+\sum_{j=1}^{d_B-1}\mu_j(c_j^*\tau_{j0}+d_j^*\tau_{jj})\;,
\end{equation}
where $\lambda$ and $\mu_j$ are real and nonnegative, the expansion coefficients $b_j$ are normalized to unity, and $|c_j|^2+|d_j|^2=1$.  The general element of ${\cal C}$ looks like
\begin{equation}\label{eq:C5}
C=\lambda^2\ket{\phi}\bra{\phi}
+\sum_{j=1}^{d_B-1}\mu_j^2\bigl(c_j\ket{0}+d_j\ket{j}\bigr)
\bigl(c_j^*\bra{0}+d_j^*\bra{j}\bigr)\;,
\end{equation}
where $\ket{\phi}$ is the state vector of Eq.~(\ref{eq:bvector}).  Thus the general element of ${\cal C}_1$ is a convex combination of an arbitrary pure state $\ket{\phi}\bra{\phi}$ and the pure states in the sum in Eq.~(\ref{eq:C5}).

We can use this example to show that $\sC_1$ (and $\sC$) are not generally convex sets.  If $\sC_1$ were a convex set, then since it contains all pure states, it would have to contain all density operators, but it is easy to find density operators not in $\sC_1$ for $d_B\ge4$.  For this purpose, we restrict attention to operators that are orthogonal to $\ket{0}$.  The general form of an element of $\sC_1$ of this sort is
\begin{equation}
C=\lambda^2\sum_{j,k=1}^{d_B}b_jb_k^*\ket{j}\bra{k}+\sum_{j=1}^{d_B}\mu_j^2\ket{j}\bra{j}\;.
\end{equation}
Let $\Pi$ be the projector onto the subspace orthogonal to $\ket{0}$, and let $\Sigma$ be any traceless, Hermitian operator that is orthogonal to $\ket{0}$, scaled so that its most negative eigenvalue is $-1$.  Then $\sigma=(\Pi+\epsilon\Sigma)/(d_B-1)$ is a density operator for $0\le\epsilon\le 1$; when $\epsilon=1$, $\sigma$ is on the boundary of the space of density operators.  For this density operator to be in $\sC_1$, we must have $\lambda^2b_jb_k^*=\epsilon\Sigma_{jk}/(d_B-1)$.  When $d_B\ge4$, we can let $\Sigma_{12}=\Sigma_{23}=-\Sigma_{13}$ be real and positive; then there is no way to
choose the phases of $b_1$, $b_2$, and $b_3$ to satisfy the condition if $\epsilon\ne0$, thus showing that $\sigma$ is not in $\sC_1$ for $0<\epsilon\le1$.

It is worth spelling this out in three dimensions ($d_B=3$), where the target unitaries on $B$ have the following matrix representations:
\begin{align}
&I=\begin{pmatrix}1&0&0\\0&1&0\\0&0&1\end{pmatrix}\;,\quad
U_1=\begin{pmatrix}1&0&0\\0&-1&0\\0&0&1\end{pmatrix}\;,\quad
U_2=\begin{pmatrix}1&0&0\\0&1&0\\0&0&-1\end{pmatrix}\;,\nonumber\\[5pt
]
&V_{01}=\begin{pmatrix}0&1&0\\1&0&0\\0&0&1\end{pmatrix}\;,\quad
V_{02}=\begin{pmatrix}0&0&1\\0&1&0\\1&0&0\end{pmatrix}\;,\quad
W_{01}=\begin{pmatrix}0&-i&0\\i&0&0\\0&0&1\end{pmatrix}\;,\quad
W_{02}=\begin{pmatrix}0&0&-i\\0&1&0\\i&0&0\end{pmatrix}\;.
\end{align}
We can see explicitly in this case that ${\cal B}$ has the form given in Eq.~(\ref{eq:B5}).  The orthocomplement $\mathcal{O}_B$, being orthogonal to all the target unitaries, is
\begin{equation}
\mathcal{O}_B=
\mbox{Span}\, \left\{\,
\begin{pmatrix}0&0&0\\0&0&1\\0&1&0\end{pmatrix},\;
\begin{pmatrix}0&0&0\\0&0&-i\\0&i&0\end{pmatrix}\,
\right\}
=\mbox{Span}\, \left\{\,
\begin{pmatrix}0&0&0\\0&0&1\\0&0&0\end{pmatrix},\;
\begin{pmatrix}0&0&0\\0&0&0\\0&1&0\end{pmatrix}\,
\right\}\;.
\end{equation}
Any operator $O_B\in\sO_B$ has all zeros in the its first row, so $\bra{0}O_B\ket{\phi}=0$ for any vector $\ket{\phi}$, implying that $S_B=\emptyset$.  This way of showing that $S_B$ is the empty set, by showing that all elements of $\sO_B$ have all zeroes in the first row, works for any value of $d_B$.
\vspace{6pt}

\noindent
\emph{Example~6.}  It is easy now to formulate an example where P holds and $R_U=2d_B$.  To do so, we consider the controlled unitary
\begin{equation}
U=\ket0\bra0\otimes I+\ket{r_1}\bra{r_1}\otimes U'_1
+\sum_{j=1}^{d_B-1}\Bigl(\ket{s_j}\bra{s_j}\otimes
V_{0j}+\ket{t_j}\bra{t_j}\otimes W'_{0j}\Bigr)\;.
\end{equation}
where the unitaries on $B$ are defined by
\begin{align}
U'_1&=I-I_{01}-Z_{01}\;,\\
V_{0j}&=I-I_{0j}+X_{0j}\;,\\
W'_{0j}&\equiv I-I_{0j}-iY_{0j}\;.
\end{align}
It is easy to convince oneself that the target unitaries are linearly independent and thus that $R_U=2d_B$.  Moreover, since
$\tau_{00}=(I-U'_1)/2$ and
$\tau_{0j}=(V_{0j}-W'_{0j})/2=(X_{0j}+iY_{0j})/2$ for
$j=1,\ldots,d_B-1$, we are in the situation discussed in Example~5, so
$S_B=\emptyset$.
\vspace{6pt}

\noindent
\emph{Example~7.}
In this example, we demonstrate that $\neg$P is possible when $R_U=d_B^2-d_B$, thus showing that condition~p2 is optimal.  We assume that system~$A$ has dimension $d_B^2-d_B$, and we use the controlled unitary
\begin{align}
\begin{split}U &=\ket0\bra0\otimes I+\sum_{j=1}^{d_B-1}\,\proj{r_j}\otimes U_j +\sum_{j=1}^{d_B-1}\,\sum_{k=j+1}^{d_B-1}\,
\Bigl(\proj{s_{j\m k}}\otimes V_{jk}+\proj{t_{j\m k}}\otimes W_{jk}\Bigr)
+\sum_{j=2}^{d_B-1}\,\proj{t_{0\m j}}\otimes W_{0j}\;,
\end{split}
\end{align}
where the indicated vectors on $A$ make up an orthonormal basis.  As in Example~5, $\sB$ is the span of the target unitaries and thus can be written as
\begin{equation}\label{eq:B7}
\sB=\mbox{Span}\bigl\{\,
\mbox{$\tau_{00}$; $\tau_{jk}$, for $j,k=1,\ldots,d_B-1$;
$Y_{0j}$, for $j=2,\ldots,d_B-1$}\,\bigr\}\;,
\end{equation}
and the orthocomplement of $\sB$ is
\begin{equation}\label{eq:OB7}
\sO_B=\mbox{Span}\,
\big\{\,\mbox{$\tau_{01}$, $\tau_{10}$; $X_{0j}$, for $j=2,\ldots,d_B-1$}\,\}\;.
\end{equation}
Letting $\ket{\phi}= \ket{0}+\ket{1}$, we clearly have $\sO_B\ket{\phi}=\sH_B$ and thus $\neg$P.  Indeed, it is easy to see that the state vectors not in $S_B$ are those that are orthogonal to $\ket0$ or to $\ket1$.
\vspace{6pt}

To introduce our last example, we consider the Weyl (or generalized Pauli) unitary operators
\begin{equation}
\label{eq:weyl}
B_{kl}=X^kZ^l\;,\quad k,l=0,\ldots,d_B-1,
\end{equation}
where $X$ and $Z$ are the shift and phase operators in the standard basis, i.e., $X\ket{j}=\ket{(j+1)\!\!\mod d_B}$ and $Z\ket{j}=e^{i2\pi j/d_B}\ket{j}$.  The Weyl operators~(\ref{eq:weyl}) are pairwise orthogonal.  The eigenstates of $X$ are the Fourier-transformed basis states
\begin{equation}
\overline{\ket k}=\frac{1}{\sqrt{d_B}}\sum_{j=0}^{d_B-1}e^{i2\pi jk/d_B}\ket j\;.
\end{equation}
We have $X\overline{\ket k}=e^{-i2\pi k/d_B}\overline{\ket k}$ and $Z\overline{\ket k}=\overline{\ket{(k+1)\!\!\mod d_B}}$.

We assume the existence of a fiducial state vector $\ket{\phi}$ such that the state vectors $\ket{\phi_{kl}}=B_{kl}\ket{\phi}$ form a SIC-POVM, i.e.,
\begin{equation}\label{eq:SICcondition}
|\braket{\phi_{kl}}{\phi_{k'l'}}|^2=\frac{1}{d_B+1}\quad\mbox{when $kl\ne k'l'$.}
\end{equation}
The existence of such a fiducial vector for all $d_B$ was conjectured by Zauner~\cite{zauner}; the conjecture has been demonstrated for $d_B=1,\ldots,15$, 19, 24, 35, and 48 and numerically up to at least $d_B=67$ (see~\cite{qiproblem} and references therein).  Notice that any vector in the SIC-POVM could be used as the fiducial vector.

Consider now the class of controlled unitaries with Schmidt decomposition
\begin{equation}
\label{eq:controlweyl}
U=\sum_{n=1}^{R_{\m U}} \proj{n} \otimes B_{n},
\end{equation}
where the operators $B_n$, indexed by the single index $n$, comprise a subset of $R_U$ elements drawn from the Weyl operators~(\ref{eq:weyl}).  The form~(\ref{eq:controlweyl}) is a Schmidt decomposition of $U$, and the Weyl operators $B_n$ are Schmidt operators on system $B$.

According to \eqref{eq:S_Bopnorm}, the fiducial state vector $\ket\phi$ is in $S_B$ if and only if
\begin{equation}\label{eq:S_Bopnormfiducial}
\biggl\|\sum_{n=1}^{R_{\m U}} B_n \proj{\phi} B^\dagger_n\biggr\|_\infty
=\biggl\| \sum_{n=1}^{R_{\m U}} \proj{\phi_n}\biggr\|_\infty
=\| \Phi \|_\infty < d_B\;.
\end{equation}
Here we define
\begin{equation}
\Phi\equiv \sum_{n=1}^{R_{\m U}} \proj{\phi_n}
\end{equation}
to be the sum of the projectors onto the subset of SIC-POVM vectors.  Our task is to bound the operator norm of $\Phi$ in terms of $R_U$.

To do so, we consider a purification of $\Phi$ into an $R_U$-dimensional system $S$:
\begin{equation}
\ket{\upphi_{SB}}\equiv\sum_{n=1}^{R_{\m U}}\ket{n}\otimes\ket{\phi_n}\;.
\end{equation}
Since $\ket{\upphi_{SB}}$ is rank one, the spectrum of $\Phi$ is the same as the spectrum of the reduced matrix of $\proj{\upphi_{SB}}$ on $S$, which is the Gram matrix of the vectors $\ket{\phi_n}$:
\begin{equation}
\tr_B(\proj{\upphi_{SB}})=\sum_{n,m=1}^{R_U}\braket{\phi_{m}}{\phi_{n}}\ket{n}\bra{m}\;.
\end{equation}
Applying the condition~(\ref{eq:SICcondition}) that the vectors $\ket{\phi_n}$ form a SIC POVM gives
\begin{align}
\tr_B(\proj{\upphi_{SB}})
&=\sum_{n=1}^{R_U}\ket{n}\bra{n}+\frac{1}{\sqrt{d_B+1}}\sum_{\substack{n,m=1,\ldots,R_U\\
n \neq m}} e^{i\theta_{nm}}\ket{n}\bra{m}\\
&=\left(1-\frac{1}{\sqrt{d_B+1}}\right)I_S+\frac{1}{\sqrt{d_B+1}}\sum_{n,m=1}^{R_U} e^{i\theta_{nm}}\ket{n}\bra{m}\;,
\label{eq:Grammatrix}
\end{align}
with $\theta_{nm}=\arg(\m \braket{\phi_n}{\phi_{m}})=-\theta_{mn}$.  Since the first term in Eq.~(\ref{eq:Grammatrix}) is proportional to the identity on $S$, we only need to bound the norm of the second term, which we do by considering an arbitrary normalized vector $\ket c=\sum_{n=1}^{R_U}c_n\,\ket{n}$:
\begin{equation}
\biggl\langle\,c\,\biggl|\biggl(\sum_{n,m=1}^{R_U} e^{i\theta_{nm}}\ket{n}\bra{m}\biggr)\biggr|\,c\,\biggr\rangle
=\sum_{n,m=1}^{R_U} e^{i\theta_{nm}}c^*_nc_{m}\leq \sum_{n,m=1}^{R_U} |c_n||c_{m}|=\biggl(\sum_{n=1}^{R_U}|c_n|\biggr)^2\leq R_U\;.
\end{equation}
Putting everything together, we obtain
\begin{equation}\label{eq:operatornormbound}
\| \Phi \|_\infty =\big\|\tr_B(\proj{\upphi_{SB}})\big\|_\infty\leq 1+\frac{R_U-1}{\sqrt{d_B+1}}\;.
\end{equation}
The right-hand-side of Eq.~(\ref{eq:operatornormbound}) is strictly smaller than $d_B$ whenever
\begin{equation}\label{eq:R_Uopbound}
R_U \leq \lceil \sqrt{d_B+1}(d_B-1) \rceil\;.
\end{equation}
Here $\lceil x \rceil$ denotes the smallest integer larger than or equal to $x$.

We conclude from the condition~\eqref{eq:S_Bopnormfiducial} that $S_B\neq\emptyset$ for $R_U \leq \lceil \sqrt{d_B+1}(d_B-1) \rceil$, regardless of the specific choice of the $R_U$ Weyl operators $B_n$ in Eq.~(\ref{eq:controlweyl}).  In particular, $S_B$ contains the fiducial vector $\ket{\phi}$.  We now examine in detail a particular case of the joint unitary~(\ref{eq:controlweyl}).
\vspace{6pt}

\noindent
\emph{Example~8.}  Consider the controlled unitary~(\ref{eq:controlweyl}) with the specific choice of $R_U=2d_B-1$ Weyl operators
\begin{equation}
\bigl\{\mbox{$I$, $Z,\ldots,Z^{d_B-1}$, $X,\ldots,X^{d_B-1}$}\bigr\}\;.
\end{equation}
For $d_B\geq 4$, condition~(\ref{eq:R_Uopbound}) is satisfied; hence the fiducial state vector is contained in $S_B$.  Moreover, the special Fourier structure of this case makes it easy to determine all the elements of the set $S_B$ for $d_B\geq 3$ (the case $d_B=2$ is equivalent to our Example~4).  To do this, we find the set of state vectors contained in $\sC_1$; $S_B$ is the complementary set.  For $d_B\geq 3$, we now show that the state vectors in $\sC_1$ are the eigenstates $\overline{\ket j}$ of $X$ and the eigenstates $\ket j$ of $Z$.

The general element of $\sB$,
\begin{equation}
B=\lambda I+\sum_{j=1}^{d_B-1}\bigl(\eta_j X^j+\zeta_j Z^j\bigr)\;,
\end{equation}
has the matrix representation in the standard basis,
\begin{equation}\label{eq:genBmatrix}
B=\begin{pmatrix}
   \overline\zeta_0&\eta_{d_B-1}&\eta_{d_B-2}&\cdots&\eta_1\vspace{2pt}\\
   \eta_1&\overline\zeta_1&\eta_{d_B-1}&\cdots&\eta_2\vspace{2pt}\\
   \eta_2&\eta_1&\overline\zeta_2&\cdots&\eta_3\vspace{2pt}\\
   \vdots&\vdots&\vdots&\ddots&\vdots\vspace{2pt}\\
   \eta_{d_B-1}&\eta_{d_B-2}&\eta_{d_B-3}&\cdots&\overline\zeta_{d_B-1}
  \end{pmatrix}\;,
\end{equation}
where
\begin{equation}
\overline\zeta_j=\lambda+\sum_{k=1}^{d_B-1}e^{i2\pi jk/d_B}\zeta_k\;.
\end{equation}
The matrix $B$ is rank one if and only if all the order-two minors (determinants of $2\times2$ submatrices) are zero.  

We first note that minor condition implies that if any of the $\eta_j$s is nonzero, then all the matrix elements of $B$---hence, all the $\eta_j$s---are nonzero.  We deal with this case first.  The condition on minors can then be restated as the requirement that the ratio of matrix elements in the same row of different columns is independent of row, and these reduce to the following conditions:
\begin{equation}\label{eq:ratios}
{\eta_1\over\vphantom{\big|}\overline\zeta_{j+1}}={\eta_2\over\eta_1}={\eta_3\over\eta_2}=\cdots
={\eta_{d_B-1}\over\eta_{d_B-2}}={\overline\zeta_j\over\eta_{d_B-1}}\;,\quad
j=0,\ldots,d_B-1.
\end{equation}
For $d_B=2$, this is a single condition, $\eta_1^2=\overline\zeta_0\overline\zeta_1=\lambda^2-\zeta_1^2$;
as mentioned above, this takes us back to Example~4, where $S_B=\emptyset$.  For $d_B\ge3$, we have immediately that all the $\overline\zeta_j$s are equal and thus that $\zeta_j=0$ for $j=1,\ldots,d_B-1$ and $\overline\zeta_j=\lambda$ for $j=0,\ldots,d_B-1$.  Now we can find the solutions of conditions~(\ref{eq:ratios}), labeled by $k=0,\ldots,d_B-1$:
\begin{equation}
\eta_j=\lambda e^{i2\pi jk/d_B}\;.
\end{equation}
This gives $B_k=d_B\lambda\,\overline{\ket k}\,\overline{\bra k}$.  We now turn to the case where all of the $\eta_j$s are zero.  In this situation, the minor condition immediately tells us that only one diagonal element of $B$ can be nonzero.  We can also deal efficiently with this case by moving to the Fourier basis, which switches the roles of $X$ and $Z$ and of the $\eta_j$s and $\zeta_j$s; then one similarly concludes that all the $\eta_j$s are zero and that $\zeta_j=\lambda e^{-i2\pi jk/d_B}$ and $\overline\zeta_j=d_B\lambda\delta_{jk}$, $k=0,\ldots,d_B-1$, giving $B_k=d_B\lambda\,\ket k\bra k$.  This shows that the pure states in $\sC_1$ are the eigenstates of $X$ and $Z$.  All other state vectors are in $S_B$.

\section{Indirect tomography}
\label{sec:tomography}

Suppose that $S_B$ is not the empty set so that, as we have learned, there are unphysical ancilla ``states'' $\sigma_B\ngeq 0$ such that the map~\eqref{eq:ancillamodel} is a proper quantum operation.  One can ask whether this happens because the unphysicality of the ancilla is ``hidden'' and irrelevant.  In the example given in the Introduction, i.e., a joint unitary $U=U_A\otimes I_B$, the ancilla does not interact with the system; the evolution the system undergoes is independent of the ``state'' of the ancilla.  We will now show that this is not always the case: a counterintuitive situation can arise, where (\romannumeral 1)~the joint unitary $U$ can lead to an operation on the system even in cases where the ``state'' of the ancilla is unphysical, i.e., $S_B\neq\emptyset$, yet (\romannumeral 2)~at the same time, $U$ allows the tomographic  reconstruction of the ``state'' of the ancilla.

By this we mean the following.  Recall Eqs.~(\ref{eq:ancillamodel}), (\ref{eq:ESchmidt}), and (\ref{eq:G}),
\begin{equation}
\sE_\sigma(\rho_A)\equiv\tr_B\bigl( U \rho_A\otimes \sigma_B U^{\dag}\bigr)
=\sum_{n,m=1}^{R_U}G_{nm}A_n\rho_A A_m^\dagger\;,
\end{equation}
where $G_{nm}\equiv(A_n|\sE_\sigma|A_m)=\tr_B(B_n\sigma_B B_m^\dagger)$, and the operators $A_n$ and $B_n$ are Schmidt operators for $U$, as in Eq.~(\ref{eq:Schmidtdecomposition}).  We can use standard process tomography to determine $\sE_\sigma$ from measurements of an informationally complete POVM on the output states obtained from an informationally complete set of input states.  We say that $U$ allows \emph{indirect tomography\/} on $B$ if this knowledge of $\sE_\sigma$ uniquely determines $\sigma_B$.  This is equivalent to saying that the matrix $G_{nm}$ determines $\sigma_B$.

To allow such indirect tomography on the ancilla state, all we need is that different operators $\sigma_B$ map to different matrices $G_{nm}$.  This map being linear, we require that $\tr_B(B_n O_B B_m^\dagger)=0$, for $n,m=1,\ldots R_U$, implies $O_B=0$ or, equivalently, that the operators $B_m^\dagger B_n$ span the space of operators on $\sH_B$.  Thus we reach our theorem on indirect tomography.
\vspace{6pt}

\noindent
{\bf Theorem. Indirect tomography.}  The unitary $U$ with Schmidt decomposition $U=\sum_{n=1}^{R_U} A_n\otimes B_n$ allows indirect tomography if and only if
\begin{equation}\dim\Bigl(\textup{Span}\bigl\{\,B_m^\dagger B_n\bigm|m,n=1,\ldots,R_U\,\}\Bigr)=d_B^{\,2}\;.
\end{equation}
\vspace{6pt}

\noindent
We want to exhibit unitaries that allow indirect tomography even if $S_B\neq\emptyset$.  We do not have to go far: Examples~7 and~8 of Sec.~\ref{sec:examples} have exactly this property. In Example~8, it is not hard to see that the set $\bigl\{B_m^\dagger B_n\,\bigm|\,m,n=1,\ldots,2d_B-1\,\}$ comprises \emph{all\/} Weyl operators (up to irrelevant phases), which means that the unitary~(\ref{eq:controlweyl}) allows indirect tomography.

By definition, when an interaction unitary $U$ allows indirect tomography there is only \emph{one\/} ``state'' $\sigma_B$ of the ancilla compatible with the evolution  $\sE_\sigma$.  Thus the situation incarnated by Examples~7 and~8 clarifies that the complete positivity of the evolution $\mathcal{E}_\sigma$ undergone by $A$, in spite of the lack of physicality of $\sigma_B$, cannot be interpreted as saying that there is some other \emph{physical\/} state $\sigma'_B\geq 0$ of the ancilla such that $\sE_\sigma=\sE_{\sigma'}$.

\section{Conclusion}
\label{sec:conclusion}

We have introduced, answered, and explored the question of what properties a joint unitary operator, acting on a system and an ancilla, must have in order that a quantum operation on the system, based on this joint unitary, requires that the ancilla state be physical.  The answer uncovers structures, not hitherto appreciated, on the space of states of the ancilla, structures clearly connected to the question of when a superoperator is a completely positive.

\section*{Acknowledgements}
  ZJ and CMC acknowledge the support of US National Science Foundation Grant Nos.~PHY-0903953 and PHY-1005540 and US Office of Naval Research Grant No.~N00014-11-1-0082. MP thanks the University of New Mexico for hospitality during the early stages of this work and acknowledges support by NSERC and CIFAR.


%

\end{document}